\documentclass[prd,preprintnumbers,nofootinbib,aps]{revtex4-1}
\usepackage[centertags]{amsmath}
\usepackage{amssymb,url}
\usepackage{epsfig,cancel} 
\usepackage{amsmath} 
\usepackage{graphicx} 
\usepackage{array,color}
\def\c{\centering}
\def\Eqn#1{Eq.\ (\ref{#1})}
\def\Eqs#1#2{Eq.\ (\ref{#1}) and (\ref{#2})}
\def\3Eqs#1#2#3{Eq.\ (\ref{#1}), (\ref{#2}) and (\ref{#3})}

\begin{document}

\title{Magic Messengers in Gauge Mediation and signal for 125 GeV 
boosted Higgs boson}


\author{Pritibhajan Byakti}
\email{pritibhajan.byakti@saha.ac.in}
\affiliation{Theory Division, Saha Institute of Nuclear Physics, 1/AF Bidhan
Nagar, Kolkata-700064, India.}
\author{Diptimoy Ghosh}
\email{diptimoyghosh@theory.tifr.res.in}
\affiliation{Tata Institute of Fundamental Research, Homi Bhabha Road,
Mumbai-400005, India.}

\preprint{TIFR/TH/12-10}


\begin{abstract}
We consider a renormalizable 
messenger sector with magic messenger fields instead of usual SU(5) complete
multiplets. We derive the soft supersymmetry breaking terms and show that the
gaugino sector can be parameterized by only two parameters. These parameters
can be chosen appropriately to obtain various patterns of gaugino masses and
different ratios among them. The sfermion sector can also be characterized 
by two independent parameters which can be adjusted to change the relative
masses of squarks and sleptons. A judicious choice of parameters also allows 
us to achieve the lightest Higgs boson mass about 125 GeV. In this paper we focus 
on a scenario where a comparatively large hierarchy exists between the U(1) and 
SU(2) gaugino mass parameters. In such a case, the lightest Higgs boson 
originating from the decay of the next-to-lightest neutralino, 
following the direct production of chargino neutralino pair, can be considerably
boosted. We show that a boosted Supersymmetric Higgs signal with a decent signal 
to background ratio can be obtained using the jet substructure technique at LHC 
with 8 TeV (14 TeV) center of mass energy and an integrated luminosity 
of about 30 fb$^{-1}$ (15 fb$^{-1}$). 
\end{abstract}
\maketitle


\section{Introduction}

The performance of the Large Hadron Collider (LHC) at CERN has been extraordinary in the 
last year and data of about 5 fb$^{-1}$ has been already collected by each of the two major 
experimental groups CMS and ATLAS. With the data already accumulated, the LHC has left the 
LEP and Tevatron results far behind in many of the search channels. But, apart from the 
recent hint \cite{ATLASHnew,CMSHnew} about a possible 124 -- 126 GeV Higgs boson, no other signal of any new physics 
(NP) has been reported yet.

The Minimally Supersymmetric Standard Model (MSSM) which is undoubtedly one of the most 
studied models beyond the Standard Model (SM) has been a benchmark for the LHC searches and 
limits on Supersymmetric particles have been shown by both the ATLAS and CMS collaborations 
\cite{susy-cms-atlas}.

Supersymmetry (SUSY) \cite{wess-bagger,Martin:1997ns} which is a beautiful theoretical idea by 
its own right has been in the spotlight among the phenomenologists for a long time because of its 
ability to tame the quadratic divergence in the scalar sector of the SM elegantly.
However if SUSY has to be realized in nature, it must be broken so that it does not contradict 
experimental observations. From a theoretical point of view, one would expect that the SUSY is 
broken spontaneously so that the underlying microscopic Lagrangian is SUSY invariant but the 
vacuum state is not. This is the same mechanism which keeps the electroweak symmetry hidden at low 
energies. There is no consensus about how SUSY must be broken at high energies but by whichever way it 
is broken, at the electroweak scale the effect of SUSY breaking can always be parameterized by 
introducing additional terms in the Lagrangian which break SUSY explicitly. But these terms should 
only include operators which do not bring back the quadratic divergences from the quantum corrections 
to the scalar masses. These are called the SUSY breaking soft terms, all of which have positive mass 
dimension. 

Unfortunately, spontaneous breaking of SUSY in a phenomenologically acceptable way has not been achieved 
till now with only tree level renormalizable interactions. This problem has been solved by breaking SUSY 
in a sector, the so called ``hidden sector'', which has either no coupling or
very small direct coupling 
to the MSSM fields. However, there 
are some mediating interactions which carry the information of SUSY breaking from the hidden sector to 
the MSSM and generate the soft terms. The gravitational interaction as a mediator of SUSY breaking has 
been the most popular scenario historically \cite{Nilles:1983ge}.
Another very popular mechanism has been the so 
called ``Gauge Mediated SUSY Breaking (GMSB)`` \cite{Dine:1981za,Dimopoulos:1981au,Dine:1981gu,
Dine:1982zb,Giudice:1998bp,Nappi:1982hm}  
scenario where the soft terms are generated through loops 
involving some messenger fields. These messenger fields couples to the SUSY breaking sector and also have 
SM gauge group quantum numbers. As gauge interactions are flavor diagonal, this mechanism also has the 
advantage of automatically being flavor-blind which is phenomenologically attractive from the points of 
view of Flavor Changing Neutral Currents (FCNC).

The General gauge mediation (GGM), a framework developed by Meade, Seiberg, and Shih \cite{Meade:2008wd} 
generalizes models of gauge mediation to accommodate arbitrary hidden sectors including those 
which are not necessarily weakly coupled. It was shown in GGM that the soft masses of MSSM gauginos
and sfermions, to leading order in the SM gauge interactions, can be expressed in terms
of the hidden sector current correlation functions. Extension of this framework
has been studied in 
\cite{Distler:2008bt,Intriligator:2008fr,Komargodski:2008ax,Carpenter:2008wi,Buican:2008ws,Kobayashi:2009rn,Dumitrescu:2010ha}. 
It was also shown in a model independent way that non-universal gaugino masses can be obtained in GGM 
without spoiling the gauge-coupling unification. By universality in gaugino
masses we mean that these 
masses at any scale are proportional to the corresponding gauge couplings at that scale. 
However, in weakly coupled GMSB scenarios with complete multiplets of SU(5) messengers the gaugino 
masses remain universal. In general messenger models \cite{Martin:1996zb,Marques:2009yu}, one can 
achieve non-universal gaugino masses but the masses of the messenger fields get constrained in order 
to maintain gauge coupling unification.

It was shown in \cite{Calibbi:2009cp} that one can achieve unification of gauge couplings even if the new 
fields added at some intermediate scales do not form complete multiplets of
SU(5) and the 
unification is independent of the scale at which these fields are added. These specially chosen fields 
were termed as ``Magic'' fields and they were used in ordinary gauge mediation in order to get non universal 
gaugino masses with extremely large hierarchy, 1:30:200 among the U(1), SU(2) and SU(3) gaugino mass 
parameters \cite{Calibbi:2009cp},
which is phenomenologically not 
interesting. Also the obtained patterns of mass ratios were not very flexible.

In this paper we generalize their idea and write the superpotential containing
all possible renormalizable terms for the messenger sector. We derive the soft
SUSY breaking terms and show that the gaugino sector can be described by two
independent parameters which can be tuned to achieve almost any ratio of gaugino
masses. This is a novel feature of our models. We also show that the sfermion 
sector in this class of models can also be parameterized 
by two parameters which can be chosen to adjust the relative masses of squarks
and sfermions. In this class of models the squark masses can be equal to the
Higgs mass parameters $m_{_{H_u}}$ and $m_{_{H_d}}$ at the messenger scale which
can lead to a small value of the Higgsino mass parameter $\mu$ and thus, a
Higgsino-like NLSP (like models of extra ordinary gauge mediation (EOGM) with
doublet triplet splitting \cite{Cheung:2007es}). Some region of the
parameter space of this model also allows for the lightest SUSY Higgs mass about
125 GeV.

We consider a phenomenologically interesting scenario where the ratio of the U(1) and SU(2) gaugino 
mass parameters at the electroweak scale is comparatively larger than the universal case 1:2. 
In this case, because of the 
comparatively larger splitting between the next-to-lightest neutralino lightest
($\chi^0_2$) and the  
neutralino ($\chi^0_1$), the Higgs boson ($h$) coming from the decay $\chi^0_2
\to \chi^0_1 ~ h $ is 
quite boosted. Motivated by this, we study the $\ell$ + $b \bar{b}$ + $P_T{\hspace{-4mm} /}$ 
~~channel to look for SUSY Higgs boson using the jet substructure technique.

The paper is organized as follows. In the next section we briefly review the 
concept of magic fields. In Sec.\ref{s:soft.terms.magic}, we describe our models
and then study the mass spectrum of gauginos and sfermions. The phenomenology of
an explicit model of this class of models is discussed in Sec. \ref{s:pheno}.
We close our discussion in Sec. \ref{conclusion} with a brief summary of our
findings.

\section{Magic fields and unification}
\label{s:magic.unification}

In this section we will review the concept of magic fields and derive some 
relations which will be useful later in this paper. It is well known that, if 
no additional matter field is added in the intermediate scale, the SM gauge couplings 
get unified in the MSSM at the scale $ M_0 \sim2.0\times 10^{16}$ GeV. 
In general, assuming that the gauge couplings unify at some scale $M_{GUT}$, 
their one loop Renormalization Group (RG) equations reads,
\begin{equation}
\label{1e:msssrge}
\alpha^{-1}_a (\mu)= -\frac{b^{a}}{2 \pi} \ln\frac{\mu}{M_{GUT}} + \alpha^{-1}
(M_{GUT}), 
\end{equation}
where the index $a = 1, 2, 3$ represent the SM gauge groups $U(1)$, $SU(2)$ and $SU(3)$ 
respectively and $b_{a}$ are the corresponding beta functions. In the MSSM we have 
$b^{a}$ = $b_0^{a}$ = \{$\frac{33}{5},1,-3$\}. 

At any value of $\mu$, \Eqn{1e:msssrge} is an equation of a straight line in 
$\alpha^{-1}$ and $b$. It directly follows that for the unification of the gauge couplings, 
$\alpha_a^{-1}$ and $b_a$ must respect the following relation, 
\begin{equation}\label{1e:gauge.coupling.ratio}
\frac{\alpha_3^{-1}-\alpha_2^{-1}}{\alpha_2^{-1}-\alpha_1^{-1}} 
= \frac {b^{3} - b^{2}}{b^{2} - b^{1}}.
\end{equation}
Even if matter fields are added at some intermediate scale, the gauge coupling unification can be 
maintained (at one loop) so far as  \Eqn{1e:gauge.coupling.ratio} is
satisfied. The fields whose 
beta functions are such that they do not spoil \Eqn{1e:gauge.coupling.ratio}
are called Magic 
fields.  

Using the one-loop RG equation of  \Eqn{1e:msssrge} in the MSSM one has,
\begin{equation}
\label{e:msssrge}
\alpha^{-1}_a(M_Z)-\alpha_b^{-1}(M_Z)= -\frac{b_0^{a}-b_0^{b}}{2\pi}
\log\frac{M_Z}{M_0}.
\end{equation}
We now add matter fields at some intermediate scales and assume that fields added at the scale 
$m_i$ contribute to the beta functions by an amount $d_i^{a}$ (dynkin index of that field). 
The gauge couplings at any scale $\mu$ can now be written as,
\begin{equation}
 \alpha^{-1}_a(\mu)=\alpha^{-1}_a(M_Z) -\frac{b_0^a}{2\pi}\log\frac{\mu}{M_Z} -
\sum_{i \mbox{ with } m_i < \mu} \frac{d_i^a}{2\pi}\log\frac{\mu}{m_i}.
\end{equation}

If we now choose
\begin{equation}
\label{e:magic.condition}
 d_i^{a}-d_i^{b}=k_i\left(b_0^{a}-b_0^{b}\right),
\end{equation}
then it is clear that \Eqn{1e:gauge.coupling.ratio} will remain intact. 
Hence we can achieve gauge coupling unification even with the presence of intermediate 
mass scales and this is independent of the values of $m_i$. But unlike complete multiplets of say, SU(5) 
the unification scale changes in this case \cite{Martin:1995wb}. 
Let us assume that the gauge coupling constants now unify at a scale $M_{\rm GUT}$. We can 
now rewrite \Eqn{e:msssrge} as,
\begin{equation}
\label{e:eogmrge}
 \alpha^{-1}_a(M_Z)-\alpha_b^{-1}(M_Z)=-\frac{b_0^{a}-b_0^{b}}{2\pi}
\log\frac{M_Z}{M_{\rm GUT}} -\sum_{i=1,\cdots,N}
\frac{d_i^{a}-d_i^{b}}{2\pi}
\log\frac{m_i}{M_{\rm GUT}}.
\end{equation}
Comparing \Eqs{e:msssrge}{e:eogmrge} we can now write the new
unification scale 
$M_{GUT}$ in terms of the unification scale $M_0$ in the MSSM, 
\begin{equation}
\label{e:new.gut}
M_{\rm GUT} = M_0
\prod_{i=1,2,\cdots,N}\left(\frac{m_i}{M_0}\right)^{\kappa_i},
\mbox{ where }\,\kappa_i=k_i/\left(1+\sum_{j=1,2,\cdots,N} k_j\right).
\end{equation}
The change in the unified value of the gauge couplings, $\delta \alpha^{-1}(M_{ GUT})$ can also also 
be written down explicitly, 
\begin{equation}
\delta \alpha^{-1}(M_{GUT})=\frac{1}{2\pi}\sum_{i=1,2,\cdots,N}
\left(d_i^{a}-\kappa_i b_0^a-\kappa_i \sum_{j=1,2,\cdots,N}
d_j^{a}\right)\log\frac{m_i}{M_0}.
\end{equation}

An example of magic fields is the combination of fields $\phi_Q, \phi_{\overline{Q}}$ and 
$\phi_G$ whose transformation properties under the SM gauge group are given by $(3,2)_\frac16$, 
$(\overline{3},2)_{-\frac16}$ and $(8,1)_0$ respectively. Here the first number in the bracket 
is the SU(3) representation and the second number refers to the SU(2) representation. The number 
outside the bracket is the Hypercharge of the multiplet. Note that the individual fields above are 
not magic fields but the combination as a whole satisfies the magic field condition of 
\Eqn{e:magic.condition}.
The set of fields mentioned above can be obtained from the SO(10) complete multiplets 
$16,\overline{16}$ and $45$ after spontaneous symmetry breaking \cite{Calibbi:2009cp}. 
\section{Mass Spectrum of the Magic Messenger models}
\label{s:soft.terms.magic}

In this section, we will derive some general properties of the magic messenger 
models  where, unlike models of ordinary gauge mediation, the superpotential
also 
contains bare mass terms allowed by symmetry. But before we
go to the details of our model, let us first discuss those models of GMSB where
non-universal gaugino masses have been obtained and 
their qualitative differences with our model.

In ordinary gauge mediation models the messenger fields are complete multiples of 
SU(5) like $5\oplus\overline{5}$. Addition of such complete multiplets changes the 
beta functions equally for all the SM gauge groups at any scale. Clearly this does not 
change the universality of the gaugino masses. Even if all possible renormalizable 
terms are considered in the superpotential, the situation does not change and 
universality is maintained. But instead of SU(5) multiplets if one uses SU(2) and 
SU(3) irreducible representations and write the most general superpotential 
(like models of EOGM with doublet triplet splitting \cite{Cheung:2007es}), then
non-universal gaugino masses 
can be  achieved only if these doublets and triplets are charged differently
under some global symmetry.
But to achieve gauge coupling unification at the same time the masses of the messenger fields should 
satisfy some constraints. In generalized messenger models, one can easily obtain 
non-universal gaugino masses but this type of models generally bypass the question of 
unification by arguing that there may be some other fields (not messenger of SUSY breaking) 
in the theory which are charged under the SM  and using these fields it is always possible 
to get successful unification. If one does not assume this then messenger masses get constrained 
again. Another generalized messenger model is also available in the literature 
\cite{Kawase:2009sx} where messenger masses are not constrained but the GUT used there 
is an anomalous U(1) GUT.

The models we are proposing are also  generalized messenger models but here the
messenger fields  are magic fields. This class of models has two main
differences with other generalized messenger models: 
(a) gauge coupling unification is independent of messenger scales and (b)
changes in beta 
functions for each added messenger satisfy the magic relation of
\Eqn{e:magic.condition}. 
The second property has non-trivial consequences as we will see in the next two
sections.

\subsection{The models}
\label{ss:models}

Our models consists of (1) $N$ pairs of magic messenger fields, each pair
consisting of 
($\phi$, $\tilde\phi$) where $\phi$ ($\tilde\phi$) transforms under some
representation 
(its conjugate representation) of the SM gauge group, and (2) a spurion field
$X$ which 
is a standard model gauge singlet but can have charges under some other symmetries like 
the R-symmetry. Note that these $\phi$ fields are in general not single irreducible 
representations;
they can be a set of fields (each transforming under an irreducible representation) which satify the magic
condition of \Eqn{e:magic.condition}. 
We assume that the field $X$ gets a vacuum expectation value (VEV) 
through some dynamics of the hidden sector: $\langle X \rangle = M +\theta^2 F$. For the sake of 
calculational simplicity we make the assumption that two different pairs of magic messengers 
either transform under the same representation of the gauge group or have no constituent fields 
having the same transformation property. This allows us to sub-divide the 
messenger field pairs into different sets (labeled by $p$) where all members of a set have the same 
representation. Note that a set may consist of either one pair of messenger fields or many pairs 
of them. We label the different members (a particular pair of magic messengers) of a set $p$ 
by the index $i_p$. For example, if a particular set $p$ has five pairs of magic messengers then 
$i_p$ runs from 1 to 5. We now write the renormalizable superpotential to be,
\begin{equation}\label{e:supot}
W = \sum_{p}\sum_{i_p,j_p}\left(\lambda_{i_pj_p}^p X +
m_{i_pj_p}^p\right)\tilde\phi_{i_p}\phi_{j_p},
\end{equation}
where the form of the matrices $\lambda^p$ and $m^p$ are determined by global (R and/or non-R) 
symmetries of the theory. Note
that, in general different constituent fields of any magic field can get
different masses and couplings. But to form a magic field, these constituent
fields need to share the same mass. We treat the magic fields ($\phi_{i_p}$'s)
`as if they are irreducible representations of some group' and write the
couplings in the superpotential because this assumption  automatically
ensures the above requirement (of equality of the masses of the constituent
fields of each the magic messenger fields). The above superpotential is
symmetric under the
interchange of 
$\phi_{i_p}$ field with $\tilde\phi_{i_p}$ and vice versa. This symmetry is
called the messenger 
parity and it helps us to get rid of the dangerous Fayet-Iliopoulos (FI) terms.
\footnote{This FI term is proportional to the VEV of the scalar component of the 
U(1) current superfield $J$. Under messenger parity $J \to -J$, because the fields 
$\tilde\phi_{i_p}$ transform under the conjugate representation of the fields 
$\phi_{i_p}$. Hence, due to messenger parity, VEV of $J$ vanishes 
\cite{Meade:2008wd,Komargodski:2008ax}.} Throughout this paper we 
will call this class of models as ``Magic Messengers in Gauge Mediation (MMGM)''.

To calculate the soft-terms, we need to write the Lagrangian in mass eigenstates 
of fermion fields and scalar fields. Using bi-unitary transformations on the 
superfields, we first make fermion mass matrices of each set diagonal and 
real: $M_f^p=\mbox{diag}(\cdots,m_{i_p},\cdots)$. In this basis the K\"{a}hler
term
will not change. The matrices $\lambda^p F$ will change but for brevity we keep the
same symbols. Messenger parity and CP conservation imply that these matrices should 
be real and symmetric. Going to the basis $\phi_{\pm p} =
\frac{1}{\sqrt{2}}(\cdots,\phi_{i_p}\pm \tilde\phi_{i_p}^*,\cdots)^T$, we can
bring sfermion mass squared matrix of any set in the block-diagonal 
form with two blocks: $M^2_{\pm p} =(M_f^p)^2\pm \lambda^p F$. Now these two
matrices can be diagonalized as $U^{p\dagger}_{\pm }M_{\pm p}^2 U^p_{\pm }$
with eigenvalues $m_{\pm i_p}^2$. We define the following combinations of $U$ matrices which 
will be used in the next section,
\begin{equation}
{\cal A}_{i_p j_p}^{p\pm }=U^{p\dagger}_{\pm  i_p j_p} U^p_{\pm
 j_p i_p}\mbox{ and } {\cal B}_{i_p j_p}=\sum_{k_p l_p}
(U^{p\dagger}_{+i_pk_p}U_{-k_pj_p})(U^{p\dagger}_{-j_pl_p}U_{+l_pi_p})\, .
\end{equation}

\subsection{Gaugino masses}
\label{s:gaugino.masses}

The expression for gaugino masses at the messenger scale can be written as, 
\begin{eqnarray}
\label{e:gaugino mass}
M_{a} = \frac{\alpha_{a}}{4\pi} \sum_p d_p^a \Lambda_p^G
\end{eqnarray}
where $d_p^a$ is the dynkin index of the magic messenger pairs in the set labeled
by $p$ corresponding to the gauge group label $a$, the superscript $G$ in $\Lambda$ refers to 
the gaugino sector, and $\Lambda^G_p$ is given by \cite{Marques:2009yu}
\begin{equation}
\label{e:diego}
\Lambda_p^G = 2 \sum_{i_p,j_p\pm}(\pm) {\cal A}^{p\pm}_{j_p i_p}
  \frac{m_{i_p} m_{\pm j_{p}}^2}{m_{\pm j_p}^2- m_{i_p}^2}
\log\left(\frac{m_{\pm j_p}}{m_{i_p}}\right)^2.
\end{equation}
Here $\Lambda_p^G$ are independent of the SM gauge group. However the presence of 
$d_p^a{}$ in the expression for gaugino masses implies that the ratio of gaugino masses 
is not equal to the ratio of $\alpha_a$'s. This means that the gaugino masses are non-universal 
in this class of models. Apparently it seems that the gaugino masses are completely independent, 
but actually this is not true. Their ratio satisfies some beautiful structure. 
The ratio of the gaugino masses can be written as, 
\begin{eqnarray}
\label{ratio}
M_1 : M_2 : M_3 &=& 1: \frac{\alpha_2}{\alpha_1}\left[1
+ \left(b_0^{(2)}-b_0^{(1)} \right) \zeta \right] : \frac{\alpha_3}{\alpha_1}
\left[1 + \left(b_0^{(3)}-b_0^{(1)} \right) \zeta \right],
\end{eqnarray}
where 
\begin{equation}
\label{zeta}
\zeta =  \frac{\sum_p k_p \Lambda_p^G}{\sum_p d_p^{(1)} \Lambda_p^G}.
\end{equation}
The value of $\zeta$ can be zero, positive or negative. All the gaugino mass 
parameters become equal if
$$\zeta=\zeta_0=-\frac{\alpha_3-\alpha_2}{\alpha_3(b_0^{(3)}-b_0^{(1)})-
\alpha_2(b_0^{(2)}-
b_0^{(1)})}=-\frac{\alpha_2-\alpha_1}{\alpha_2\left(b_0^{(2)}-b_0^{(1)}\right)},
$$
where in the last step we used \Eqn{1e:gauge.coupling.ratio}.
Any arbitrary $\zeta$ can always be written as
\begin{eqnarray}\label{e:zeta.tilde}
 \zeta=\zeta_0 + \tilde\zeta.
\end{eqnarray}
 In that case \Eqn{ratio} takes
the form,
\begin{eqnarray}
\label{ratio2}
M_1 : M_2 : M_3 &=& 1 : 1 + \frac{\alpha_2}{\alpha_1} \left(b_0^{(2)}-b_0^{(1)}
\right) \tilde\zeta : 1 + \frac{\alpha_3}{\alpha_1} \left(b_0^{(3)}-b_0^{(1)}
\right) \tilde\zeta \, , \nonumber\\
&=& 1 : 1 -\frac{28}{5}\frac{\alpha_2}{\alpha_1}\tilde\zeta :  1
-\frac{48}{5}\frac{\alpha_3}{\alpha_1}\tilde\zeta.
\end{eqnarray}

It is now clear from \Eqn{ratio2} that various ratios among the gaugino
masses can be obtained,
(a) for negative $\tilde\zeta$, one gets normal hierarchy, (b) for $\tilde\zeta=0$, all
the gaugino masses are equal, and, (c) for non-zero positive 
$\zeta\leq\frac{5\alpha_1}{48\alpha_3}$, one gets inverted hierarchy. For 
$\zeta> \frac{5\alpha_1}{48\alpha_3}$, some more patterns can be achieved. 

So to conclude this section, the gaugino sector of this class of models can be parametrized by two 
free parameters: one of them can be taken as the mass of the U(1) gaugino and 
the other one is the $\zeta$ (or $\tilde\zeta$) parameter.
These two parameters can be tuned to get phenomenologically 
interesting ratios among the gaugino masses.

\subsection{Sfermion masses}\label{ss:scalar}

The sfermion masses are obtained at the two loop level in GMSB. In GGM, these are 
parameterized by three parameters $A_a$ (a =1, 2, 3) for three gauge groups U(1), SU(2) and SU(3). 
In terms of these three parameters the expression for the sfermion masses \cite{Meade:2008wd} can be 
written as follows,
\begin{eqnarray}
\label{mass-sfermions}
m^2_{\tilde f} &=& \sum_a g^4 c_2(f,a) A_a,
\end{eqnarray}
where $c_2(f,a)$ is the quadratic Casimir of the representation $f$ of the gauge group $a$. 
In our class of models the expression for $A_a$ turns out to be
\begin{eqnarray}
\label{A-parameter}
 A_a& =& 2\frac{1}{(16\pi^2)^2}\sum_p d_p^a\Lambda_p^S
\end{eqnarray}
where $\Lambda_p^S$ is given by \cite{Marques:2009yu}, 
\begin{eqnarray}
 \Lambda_p^S &= &2 \sum_{i_pj_p\pm} m^2_{\pm i_p}\left[ {\cal A}_{i_p j_p}^{p\pm
}    \log\left(\frac{m_{\pm i_p}^2}{m_{j_p}^2}\right) - 2\,
{\cal A}_{i_p j_p}^{p\pm }  Li_2\left( 1-\frac{m_{j_p}^2}{m_{\pm
i_p}^2}\right)\right.\nonumber\\ &&\left. + \frac12 {\cal B}_{i_p j_p}
Li_2\left( 1-\frac{m_{\mp
j_p}^2}{m_{\pm i_p}^2} \right)
\right].
\end{eqnarray}
The quantities $\Lambda^S_p$ and $A_a$ are required to be strictly positive in order to have 
positive sfermion masses. 

Ratio of the $A_a$ parameters can be written as, 
\begin{equation}
\label{A-ratio}
A_1: A_2 : A_3 = 1: 1-\frac{28}{5}\eta:1- \frac{48}{5}\eta ,
\end{equation}
where 
\begin{equation}
\label{eta}
\eta= \frac{\sum_p k_p \Lambda^S_p}{\sum_p d_p^{(1)}\Lambda_p^S}.
\end{equation}
Note that $\eta$ cannot be greater than $\eta_0 = \frac{5}{48}$. For
$\eta\sim \eta_0$, one has 
$A_1> A_2 \gg A_3$. In this limit, from \Eqn{mass-sfermions} one can see that
soft SUSY 
breaking squark and Higgs masses will be 
approximately same (universal) at the messenger scale. This would in turn render the value of 
$m_{_{H_u}}$ at weak scale insensitive to the universal scalar masses in the ultraviolet. 
This can be understood as a consequence of a ``focus point`` in the RG behavior of 
$m_{_{H_u}}$ \cite{Feng:1999mn,Feng:1999zg}. In this case it is possible to have a small 
$\mu$ (supersymmetric mass term for the Higgs) \cite{Agashe:1997kn,Agashe:1999ct} and thus, 
a Higgsino like NLSP. This can lead to very distinct phenomenology \cite{Ruderman:2011vv,Kats:2011qh}.

\section{An explicit model of MMGM}

In this section we construct an explicit model of MMGM. We take two sets of messengers, 
the first set constists of two pairs of SU(5) $5+\overline{5}$ fields: $\phi_i,\tilde\phi_i$ 
with $i=1,2$, and the second set contains a magic field: $\phi_3=\phi_Q + \phi_{\overline{Q}} +
\phi_G $. 
Note that one can write down the required terms for gauge mediation 
(as given in \Eqn{e:supot}) using only $\phi_3$ and without invoking its 
conjugate field $\tilde\phi_3$. Moreover, absence of the field $\tilde\phi_3$ does 
not break the messenger parity because the superpotial remains invariant under the 
interchange of $\phi_Q$ by $\phi_{\overline{Q}}$ and vice versa. This is why we 
do not include the $\tilde\phi_3$ field in the second set just to make it more 
economical. 
In the second set, we will denote the fermion mass by $m_3$ and the eigenvalues
of the scalar mass 
squared matrix by $m_3^2\pm d$. For the first set, in the diagonal basis of the fermion mass matrix 
$M_f=\mbox{diag} (m_1, m_2)$, the mass squared matrix of the scalars will look like,
\begin{equation}
\label{1mass-square}
M^2_{\pm}= \left(
\begin{array}{cc}
  m_1^2 &0 \\ 
0 & m_2^2 
\end{array}  \right) 
\pm 
\left(\begin{array}{cc}
a & b \\
b& c
\end{array}
 \right).
\end{equation}
Clearly the eigenvalues and the diagonalizing matrices of $M^2_\pm$ will involve the quantities 
$a$, $b$, $c$ and hence, the four parameters $M_1, \tilde\zeta, A_1$ and $\eta$ will also depend 
on them. 
Just to give few examples, in Table-\ref{t:atoA} we choose six benchmark points
\footnote{In our models, gravitino mass is in the range 17 GeV
to 20 GeV. If we assume gravitino to be the LSP then Big Bang Neucleosynthesis will
be problematic because of large neutralino lifetime. This problem can be
easily solved in the axino dark matter models where life time of the lightest
neutralino can be less than .1 sec\cite{Asaka:2000ew,Covi:1999ty,Covi:2001nw}.}
 and show the values of $a, b, c,
d$ and the 
corresponding numerical values for $M_1, \tilde\zeta, A_1$ and $\eta$.
\begin{table}[h]
\begin{center}
\begin{tabular}{|c|c|c|c|c|c|c|c|c|c|c|c|}
\hline
 Benchmark    & $a$ & $b$ & $c$ & $d$ &  $M_1$  & \LARGE{$_{\tilde\zeta}$} & \c
$A_1$  & $\eta$\\
points & & & & & (GeV)& & ($10^{8}$ \phantom{q}GeV$^2$) & ($10^{-5}$) \\
\hline
\hline
1  & 0.85 & 80  &0.85 & 0.60   & 97.304 & -0.1914 & 2.0534 & -1.4061\\
\hline
2 & 0.80 & 70  &0.80 & 0.90   & 95.165 & -0.2794 & 1.5722 & -4.1320\\
\hline
3 & 0.80 & 70 &0.80 & 1.2 & 98.373 & -0.3527 & 1.5722 & -7.3458\\
\hline
4 &0.80 & 70  &0.80 & 1.4  & 100.51& -0.3989 & 1.5722& -9.9983\\
\hline
5 &0.80 & 70 &0.80 & 1.6  & 102.65 & -0.4432 & 1.5723 & -13.059\\
\hline
6 &0.80 & 70 &0.80 & 1.8  & 104.79 & -0.4858 & 1.5723 & -16.527\\
\hline
\end{tabular}
\end{center}
\caption{The values of $a, b, c$ and $d$ parameters for six benchmark points
 and the 
corresponding numerical values for the quantities $M_1$, $\tilde\zeta$, $A_1$ and $\eta$. 
The values of $m_1$, $m_2$ and $m_3$ are $1.0\times 10^{14}$ GeV.  The 
parameters $a, b, c$  and $d$ are given in units of $10^{18}$ GeV$^2$. }
\label{t:atoA}
\end{table}

It is well known that among the various models of GMSB, the mass ratio between the sfermions 
and gauginos is the lowest in case of ordinary gauge mediation models 
\cite{Dumitrescu:2010ha} 
(which can be obtained by setting $b = 0$ in \Eqn{1mass-square}). 
In order to increase this ratio, $b$ should be made large compared to $a$ and $c$. 
The parameter $d$ is related to the magic part of the model. So changing $d$, one can increase 
or decrease the splitting among the gaugino masses. Note that, we got
$\tilde\zeta\sim 10^{-1}$ whereas $\eta\sim 10^{-5}$. So gaugino
masses are highly hierarchical (see Table II) but $A_a$ parameters are not. 
To understand this one should note that the SU(5) part of this model is not an 
OGM model but an EOGM model, and $b$ is hundred times larger than $a$, $c$ and $d$. 
From Eqn. (B.6) and (B.8) of \cite{Dumitrescu:2010ha}, one can see that the 
off-diagonal element $b$ has no effect on the expression for gaugino masses whereas 
it has dominant contribution on the sfermion masses. Using the fact that 
$\frac{F}{M^2}\sim10^{-8}$, the expression for the $\eta$ parameter can be
approximated as: $\eta\sim-\frac14\frac{d^2}{b^2}$. This explains why $\eta$ is
small in this model.

\subsection{Phenomenology: boosted higgs signal}\label{s:pheno}
\label{pheno}

There are many models of supersymmetry breaking where the soft supersymmetry
breaking 
gaugino mass terms M$_1$, M$_2$ and M$_3$ meet to a common value m$_{1/2}$ at
the GUT 
scale. Now the one-loop Renormalization Group equations for the three gaugino
mass 
parameters in the MSSM are determined by the same quantities $\rm b_{\rm a}^0$
(a=1, 2, 3) 
which also control the RG running of the three gauge couplings. It then
immediately 
follows that each of the three ratios $\frac{\rm M_a}{\rm g^2_a}$ is one-loop RG
invariant. 
In models of gaugino mass 
unification (for example, models with minimal supergravity or gauge-mediated
boundary 
conditions) this leads to an interesting relation, M$_1$:M$_2$:M$_3$ = 1:2:7
approximately 
at the TeV scale (modulo two loop corrections and unknown threshold
corrections).
Now, if 
the supersymmetric Higgsino mass parameter 
$\mu$ $\gg$ M$_1$, M$_2$ then the physical mass eigenstates consist of a
``bino-like'' 
lightest neutralino $\tilde\chi_1^0$ and a ``wino-like'' next-to-lighest
neutralino 
$\tilde\chi_2^0$ and lightest chargino $\tilde\chi_1^\pm$. In this case one has
to a very 
good approximation 
$m_{\tilde g}$ $\sim$ M$_3$, $m_{\tilde\chi_2^0}, m_{\tilde\chi_1^\pm}$ $\sim$
M$_2$ and 
$m_{\tilde\chi_1^0}$ $\sim$ M$_1$. With the increase of the lower bound on
squark and 
gluino masses by the LHC data, their production cross section has been pushed to
quite 
low values ($\sim$ few fb). On the other hand the electro weak gaiginos can be 
sufficiently lighter ($\tilde\chi_1^\pm,\tilde\chi_2^0$ $\sim$ 250 is still
allowed even 
if universality is assumed). Hence the pair production cross-section of the
light
electroweak gauginos dominate the SUSY production cross-section. 

In our model, as we have already seen, in general the gaugino mass parameters
can be 
arranged to have any ratio among themselves. In particular, we consider the
case where 
M$_1$, M$_2$ and M$_3$ are much more hierarchical than the ratio 1:2:7. In this
case 
$\tilde\chi_1^\pm$ and $\tilde\chi_2^0$ even lighter than 250 GeV are still
allowed.
In this section we 
consider the production of $\tilde\chi_1^\pm,\tilde\chi_2^0$ $\sim$ and their
subsequent 
decays, $\tilde\chi_1^\pm \to \tilde\chi_1^0 W$ and $\tilde\chi_2^0 \to
\tilde\chi_1^0 h$. 
The prospect of this channel at 8 TeV LHC in case of the mSUGRA model has
been studied 
in detail in \cite{Ghosh:2012mc}. It was concluded that an integrated luminosity
of 
100 fb$^{-1}$ 
will be needed for a good signal to background ratio. Note that, unlike mSUGRA
in our 
model the lightest neutralino $\chi_1^0$ can be considerably lighter than
$\chi_2^0$ and 
hence, the lightest Higgs boson from the decay of $\chi_2^0$ can be quite
boosted. 

In Table-\ref{table1} we show a few benchmark points which will be used for a
detail signal 
anlysis. We choose the parameters in our model so as to get different ratios of
gaugino 
masses along with the lightest Higgs boson mass consistent with the recent hints
of Higgs 
signal by the CMS \cite{Chatrchyan:2012tx} and ATLAS \cite{:2012si}
collaborations. Note that, in our case the branching ratio for the decay 
$\tilde\chi_2^0 \to \tilde\chi_1^0 h$ is very large ($> 95\%$). For 
related discussions on the parameter dependence on this branching ratio and interplay 
with other decay modes see the reference \cite{Gori:2011hj}.

To generate the mass spectrum of the SUSY particles we have used the package 
SuSpect \cite{Djouadi:2002ze}. SUSYHIT \cite{Djouadi:2006bz} has been used to calculate the 
corresponding branching ratios. Note that the gluinos are quite heavy for all the benchmark points and consequently their production cross-section is extremely small. For example, for the benchmark point-1 the gluino pair production cros-section is about 4 fb (LO) for 8 TeV center of mass energy and it is even smaller for the other benchmark points.

\begin{table}[h!]
\begin{center}
 \begin{tabular}{|p{1.8 cm}|m{.9 cm}|p{1.2cm}|p{1.2cm}|p{1.2cm}
|p{1.2cm}|p{1.2cm}|p{1.2cm}|c|c|p{1.cm}|}
\hline
\c Benchmark points 
&\c $\rm tan \beta$ & \c $m_{\tilde\chi_1^\pm}$ (GeV) & \c
$m_{\tilde\chi_2^0}$ (GeV) & \c $m_{\tilde\chi_1^0}$ (GeV) & \c
 $m_h$ (GeV) & $\hspace{3mm} m_{\tilde g}$ \hspace{3mm} (GeV) & 
$\hspace{3mm}\mu $ \hspace{3mm} (GeV) & BR$(\chi_2^0 \to \chi_1^0 h )$ &
$\frac{\delta M_Z^2}{M_Z^2}(B\mu)$ & $\frac{\delta M_Z^2}{M_Z^2}(\mu)$ $(10^4)$
\\
\hline
\hline
\c 1 &  \c 20 & \c 203.0 & \c 203.0 & \c 50.9 & \c
124.8 & \hspace{3mm} 1019& \hspace{3mm} 7935 & 98.908 & 66.25 & 1.53
\\
\hline
\c 2 & \c 25 & \c 248.1 & \c 248.1 & \c 49.0 & \c
124.6 &\hspace{3mm} 1239 & \hspace{3mm} 7003 & 96.883 & 30.83 & 1.19\\
\hline
\c 3 & \c 30 & \c 299.4 & \c 299.4 & \c 50.2 & \c
124.8 &\hspace{3mm} 1482 & \hspace{3mm} 7024 & 95.221 & 19.74 & 1.19\\
\hline
\c 4  & \c 20 & \c 332.3 & \c 332.3 & \c 50.6 & \c 124.8 &\hspace{3mm} 1641
&\hspace{3mm} 7055 & 96.076 & 52.64 & 1.21\\
\hline
\c 5 &  \c 20 & \c 366.1 & \c 366.1 & \c 51.3 & \c 124.9
&\hspace{3mm} 1797 &\hspace{3mm} 7070 & 95.621 & 52.92 & 1.21\\
\hline
\c 6 &  \c 20 & \c 399.7 & \c 399.7 & \c 51.9 & \c 124.9 &\hspace{3mm} 1950
&\hspace{3mm} 7085 & 96.343 & 53.19 &1.22\\
\hline
\end{tabular}
\end{center}
\caption{Masses of the lightest neutralino, next-to-lightest neutralino,
lightest chargino and lightest CP even neutral Higgs for the six benchmark points. 
All other SUSY particles have masses in the multi-TeV range
($\sim$ 7 TeV to 15 TeV). Definition of fine tuning parameters
 are $\frac{\delta M_Z^2}{M_Z^2}(\mu)=\frac{2\mu^2}{M_Z^2} \left[ 1+ t_\beta
\frac{4 \tan^2\beta (\bar{m}_1^2-
\bar{m}_2^2) }{ (\bar{m}_1^2-\bar{m}_2^2)t_\beta-M_Z^2} \right]\frac{\delta
\mu^2}{\mu^2}$ and $\frac{\delta M_Z^2}{M_Z^2}(B\mu)= 4 t_\beta \, \tan^2\beta 
\, \frac{
\bar{m}_1^2-\bar{m}_2^2 }{ M_Z^2 (\tan^2\beta -1)^2}$ where
$t_\beta=(\tan^2\beta+1)/(\tan^2\beta-1)$. Hence fine-tuning is large. }
\label{table1}
\end{table}

\begin{figure}[h!]
\begin{center}
\includegraphics[]{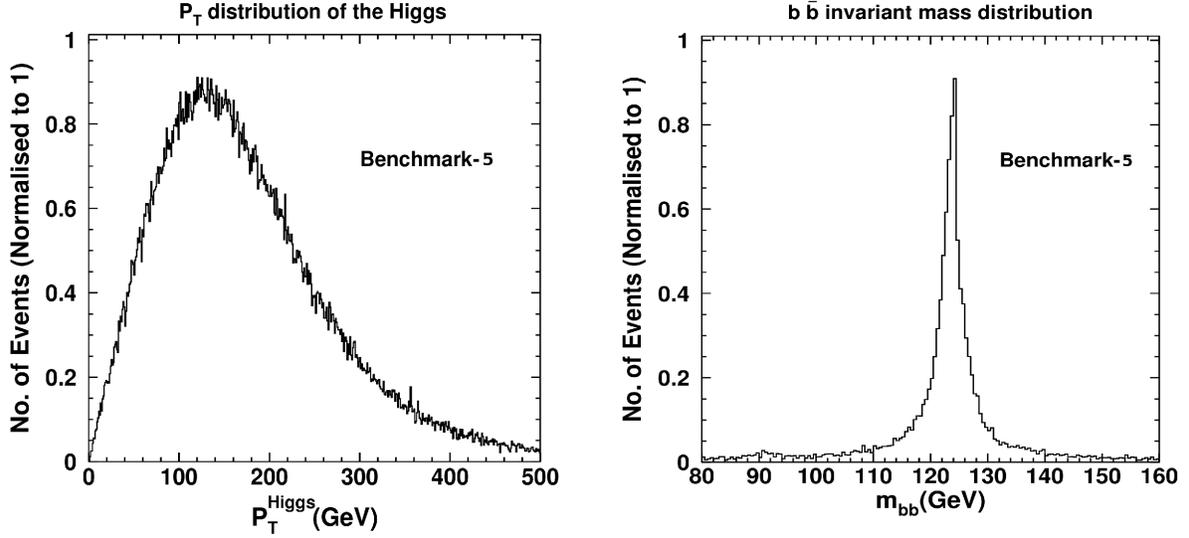}
\caption{Left panel : The transverse momentum distribution of the Higgs boson
from the decay 
$ \chi_2^0 \to \chi_1^0 \; h$ following the direct production of ($\chi_1^\pm,
\chi_2^0$) 
at 8 TeV LHC for our SUSY benchmark point 5. The y-axis has been normalized to
1. 
\newline
Right panel: The invariant mass distribution of the reconstructed Higgs boson
using the jet 
substructure algorithm, as described in the text, for our SUSY benchmark point
5. 
The y-axis has been normalized to 1. The figures have been generated using the
CERN 
package PAW \cite{cern:paw}.}
\label{fig1}
\end{center}
\end{figure}

In Fig.\ref{fig1} we show the transverse momentum distribution of the Higgs in
the decay 
$\tilde\chi_2^0 \to \tilde\chi_1^0 h$ following the direct production of 
$\tilde\chi_1^\pm,\tilde\chi_2^0$ at the LHC with 8 TeV center of mass energy.
It can be 
observed that a large fraction of the Higgs bosons has transverse momentum
greater than 
100 GeV. This allows us to use the jet substructure technique to look for Higgs
in the 
decays of directly produced electroweak gauginos. 

The use of jet substructure for the reconstruction of hadronic decays of 
boosted W, Z, Higgs bosons and top quarks has received considerable 
attention in recent years \cite{Kribs:2010hp,Abdesselam:2010pt,Altheimer:2012mn}. A study of jet substructure in the context of a 
search for a heavy Higgs boson 
decaying to W W was first carried out in Ref.~\cite{Seymour:1993mx}. 
More recently, Butterworth, Davison, Rubin and
Salam(BDRS)\cite{Butterworth:2008iy} 
studied the case of a light Higgs boson ($\rm m_{\rm H}$ $\sim$ 120 GeV)
produced 
in association with an electroweak gauge boson. The leptonic decay of the
associated 
vector boson provides an efficient trigger for these events. The BDRS algorithm 
involves a technique using the mass-drop and the filtering to transform the 
high-p$_{_{\rm T}}$ WH, ZH(H $\to$ b $\bar{\rm b}$) channel into one of the
best 
channels for discovery of Standard Model Higgs with small mass at the LHC. 

In this section we adopt the BDRS method for tagging hadronically decaying Higgs
boson. 
We describe below the exact procedure adopted in our analysis in order to
implement 
this along with our other selection cuts. 

We first cluster all the stable final state particles(excluding leptons,
neutrinos and 
neutralinos) into ``fat-jets'' using the Cambridge-Aachen algorithm (CA
algorithm) 
\cite{Dokshitzer:1997in,Wobisch:1998wt} as implemented by the Fastjet 
package \cite{Cacciari:2005hq} with R parameter of 1.2. 
We then select the jets with transverse momentum p$_{_{\rm T}} > $ 100 GeV and 
pseudo rapidity $|\eta|$ $<$ 2.5. 
We now perform the jet substructure analysis on the hardest jet following the
BDRS 
prescription. Here we closely follow the discussion of
Ref.~\cite{Butterworth:2008iy}, 
mentioning our specific choices of parameters as and when the occasion arises.
The first step is to select a jet-j and apply the following procedure:
\begin{itemize}
\item[1.] Undo the last clustering step of the jet-j to get two subjets. Label
the two 
subjets j$_1$ and j$_2$ so that m$_{\rm j_1}$ $>$ m$_{\rm j_2}$. 
(Remember that at each step during the CA clustering the masses of the
proto-jets 
to be combined are recorded.)

\item[2.] Define $\mu$= $\dfrac{\rm m_{\rm{j}_1}}{\rm m_{\rm j}}$,  
y$_{\rm cut}$= $\dfrac{\rm min(p^2_{\rm T \, \rm j_1} \;,\;\rm p^2_{\rm T \, 
\rm j_2})}{\rm m^2_{\rm j}}$ $\Delta \rm R^2_{\rm j_1\; , \; \rm j_2}$. Here 
$\Delta \rm R_{\rm j_1\; , \; \rm j_2}$ is the separation between the two 
subjets j$_1$ and j$_2$ in the pseudo rapidity($\eta$)- azimuthal angle($\phi$)
plane. 
If $\mu$ $<$ $\mu_{\rm cut}$ (significant mass drop) and y $>$ y$_{\rm cut}$ 
(the splitting of the hard jet-j into two 
subjets j$_1$ and j$_2$ is not too asymmetric) then go to step 4. 
We use $\mu_{\rm cut}$ =0.4 and y$_{\rm cut}$=0.1 in our analysis.

\item[3.] Otherwise redefine $\rm j = \rm j_1$ and go back to step 1. 

\item[4.] Take the constituents of the mother jet-j  and recluster them with the
Cambridge-Aachen 
algorithm with an R-parameter of R$_{\rm filt}$=min($\frac{\Delta R_{\rm j_1\, ,
\, 
\rm j_2}}{2}$\,,\,0.4). 
Construct n new subjets $\rm j^{\rm filt}_1$, $\rm j^{\rm filt}_2$, $\rm j^{\rm
filt}_3$, ..... 
$\rm j^{\rm filt}_n$ ordered in descending $\rm p_{\rm T}$. 
This step is supposed to reduce the degradation of resolution on jets caused by
underlying events.

\item[5.] Require the two hardest of the subjets $\rm j^{ \rm filt}_1$ ......
$\rm j^{\rm filt}_n$ 
to have b tags.

\item[6.] Define $\rm j^{\rm higgs} = \displaystyle\sum\limits_{i=1}^{\rm
min(n,3)} \rm j^{\rm filt}_i$. 
This step captures the dominant $\mathcal O (\alpha_{\rm s} )$ radiation from
the Higgs decay, 
while eliminating much of the contamination from underlying
events\cite{Butterworth:2008iy}.

\end{itemize}

The parameters involved in this method can be, in principle, optimized event by
event 
\cite{Kaplan:2008ie}. In our analysis, these parameters have been set to fixed
values as 
already mentioned above. In Fig.~\ref{fig1} we show the distribution of the 
reconstructed Higgs mass using the above BDRS prescription for our SUSY
benchmark point-5. 
For the other benchmark points the distributions are qualitatively same and we
do not 
show them here. 

\begin{table}[h!]
\hspace*{-5mm} \begin{tabular}{|p{2.1cm}|p{1.8cm}|p{1.5cm}|p{0.9cm}|p{1.cm}| 
p{1.15cm}|p{1.15cm}|p{1.cm} | m{1.8cm} | c }
\hline
\multicolumn{3}{|c|}{} & \multicolumn{5}{c}{Number of Events after individual
cuts}& 
\multicolumn{2}{|c|}{}\\
\hline
\c Process & \c Crosssection (LO) & \c Simulated Events & Cut-I & Cut-II &
Cut-III & 
Cut-IV & Cut-V & \c Crosssection after Cut-V & 
\multicolumn{1}{m{2cm}|}{$\dfrac{\rm S}{\sqrt{\rm B}}$(30 fb$^{-1}$)}
\\
\hline
\multicolumn{9}{|c|}{Signal} \\
\hline
Benchmark-1&\c 533 fb&\c 50K&\c 681 &\c 178&\c 71&\c 47&\c 28&
\hspace{3mm} 0.21 fb&
\multicolumn{1}{c|}{$\approx 4 $ }\\
\hline
Benchmark-2&\c 241 fb &\c 50K&\c 1220&\c 311&\c 150&\c 106&\c 84&
\hspace{3mm} 0.27 fb&
\multicolumn{1}{c|}{$\approx 5.1 $ }\\
\hline
Benchmark-3&\c 108 fb&\c 50K&\c 1729&\c 475&\c 287&\c 201&\c 177&
\hspace{3mm} 0.27 fb &
\multicolumn{1}{c|}{$\approx 5.1 $ }\\
\hline
Benchmark-4&\c 71 fb&\c 50K&\c 2112&\c 552&\c 356&\c 270&\c 246&
\hspace{3mm} 0.23 fb &
\multicolumn{1}{c|}{$\approx 4.4\phantom{.0}$ }\\
\hline
Benchmark-5&\c 45 fb&\c 50K&\c 2697&\c 788&\c 522&\c 380&\c 341&
\hspace{3mm}  0.21 fb&
\multicolumn{1}{c|}{$\approx 4 $ }\\
\hline
Benchmark-6&\c 30 fb&\c 50K&\c 3092&\c904 &\c 655&\c488&\c449&
\hspace{3mm}   0.21 fb &
\multicolumn{1}{c|}{$\approx 4 $ }\\
\hline
\multicolumn{9}{|c|}{Background} \\
\cline{1-9}
\c t $\bar{\rm t}$(0-100)& \c 48.3 pb&\c 40L&\c 1126 &\c 274 &\c 29 &\c 3 & \c 0  & 
  \\
\cline{1-9}
\c t $\bar{\rm t}$(100-200)&\c 36.6 pb&\c30L &\c 976 &\c 234 & \c 26 & \c 6 & \c
1  &   0.0119 fb \\
\cline{1-9}
\c t $\bar{\rm t}$(200-300)&\c 7.8 pb&\c 5L&\c 157&\c 40&\c 6&\c 3&\c 2&  0.0306 fb \\
\cline{1-9}
\c t $\bar{\rm t}$(300-500)& \c 1.8 pb &\c 1.5L & \c 24 & \c 5 & \c 1 &\c  0 &\c
0 &    \\
\cline{1-9}
\c t $\bar{\rm t}$(500-$\infty$)& \c 134 fb & \c 10K & \c 0 & \c 0 & \c 0 &\c 0
&\c 0&  \\
\cline{1-9}
W($\to l\,\nu_{_l}$) b $\bar{\rm b}$ $\phantom{ss} l = e, \mu, \tau$
& \c 3 pb & \c 201308 & \c 64 &\c  35 & \c 3 & \c 2 & \c 0  &    \\
\cline{1-9}
\c W h               & \c 549 fb & \c 50K &\c 401 &\c 125 & \c 9 & \c 5 &\c 2 & 0.0108
\\
\cline{1-9}
\c W Z               & \c 13 pb  & \c 8L  & \c 11  &\c 3  &\c 0  &\c 0  &\c 0   &
   \\
\cline{1-9}
\c Z h               & \c 296 fb & \c 20K & \c 167 & \c 4 &\c 0  &\c 0  &\c 0  
&    \\
\cline{1-9}
Z($\to l^+ l^-$)b$\bar{\rm b}$ $\phantom{ss} l = e, \mu, \tau$
& \c 2 pb &\c 126581  &\c 26  &\c 5 &\c  0  &\c  0 & \c 0  &    \\
\cline{1-9} 
t($\to e\,\nu_{e}\,\rm b$) b & \c 308 fb    & \c 36817  & \c 6   &\c 5 &\c 0  &
\c 0 & \c 0  &    \\
\cline{1-9}
\c t b W & \c 18.7 pb  & \c 597812 & \c 156 & \c 43 & \c 4 & \c 2 & \c 2 &
0.0306 fb \\
\cline{1-9}
\c Total Background& \multicolumn{7}{c|}{} &  0.084 fb     \\
\cline{1-9}
\end{tabular}
\caption[]{Event summary for the signal and the backgrounds after individual
cuts as 
described in the text for LHC8. In the second column, the Leading Order (LO)
cross-sections 
have been obtained using PYTHIA and ALPGEN. While calculating the final crosssection 
after Cut-V, the NLO cross-sections for signal(from PROSPINO) and appropriate K
factors 
(whenever available) for the backgrounds, as mentioned in the text, have been used. 
The b-tagging efficiency has 
also been multiplied. K and L in the third column stand for $10^3$ and
$10^5$ respectively. Note that the number of simulated events is always more than the 
expected number of events at LHC (at 30 fb$^{-1}$) for both the signal and backgrounds.}
\label{table2}
\end{table}
\begin{table}[h!]
\hspace*{-5mm} \begin{tabular}{|p{2.1cm}|p{1.8cm}|p{1.5cm}|p{0.9cm}|p{1.cm}| 
p{1.15cm}|p{1.15cm}|p{1.cm} | m{1.8cm} |}
\hline
\multicolumn{3}{|c|}{} & \multicolumn{5}{c}{Number of Events after individual
cuts}& 
\multicolumn{1}{|c|}{}\\
\hline
\c Process & \c Crosssection (LO) & \c Simulated Events & Cut-I & Cut-II &
Cut-III & 
Cut-IV & Cut-V & Crosssection after Cut-V \\
\hline
\multicolumn{9}{|c|}{Signal} \\
\hline
Benchmark-1&\c 1.4 pb&\c 50K&\c 769 &\c 186&\c 78&\c 54&\c 29&
\hspace{3mm} 0.53 fb  \\
\hline
Benchmark-2&\c 681 fb &\c 50K&\c 1282&\c 327&\c 156&\c 119&\c 92 &
\hspace{3mm} 0.73 fb \\
\hline
Benchmark-3&\c 332 fb&\c 50K&\c 1857&\c 465&\c 288&\c 198&\c 169&
\hspace{3mm} 0.73 fb \\
\hline
Benchmark-4&\c 226 fb&\c 50K&\c 2363&\c 626&\c 392&\c 274&\c 238&
\hspace{3mm} 0.73 fb \\
\hline
Benchmark-5&\c 155 fb&\c 50K&\c 2792&\c 746&\c 499&\c 359&\c 321&
\hspace{3mm} 0.67 fb \\
\hline
Benchmark-6&\c 105 fb&\c 50K&\c 3125&\c843 &\c 592&\c416&\c373&
\hspace{3mm} 0.53 fb\\
\hline
\multicolumn{9}{|c|}{Background} \\
\cline{1-9}
\c t $\bar{\rm t}$(50-$\infty$)& \c 335 pb&\c 100L&\c 3467 &\c 913 &\c 128 &\c 11  
& \c 3  & 0.098 fb \\
\cline{1-9}
W($\to l\,\nu_{_l}$) b $\bar{\rm b}$ $\phantom{ss} l = e, \mu, \tau$
& \c 5.45 pb & \c 429871 & \c 148 &\c 86 & \c 7 & \c 6 & \c 0  &    \\
\cline{1-9}
\c W h               & \c 1.24 pb & \c 50K &\c 446 &\c 108 & \c 7 & \c 3 &\c 1 
& 0.012 fb
\\
\cline{1-9}
\c W Z               & \c 29 pb  & \c 8L  & \c 20  &\c 5  &\c 0  &\c 0  &\c    &
   \\
\cline{1-9}
\c Z h               & \c 674 fb & \c 50K & \c 449 & \c 6 &\c 0  &\c 0  &\c 0  
&    \\
\cline{1-9}
Z($\to \ell^+ \ell^-$)b$\bar{\rm b}$ $\phantom{ss} \ell = e, \mu, \tau$
& \c 7.4 pb &\c 607607  &\c 116  &\c 23 &\c  5  &\c  4 & \c 0  &    \\
\cline{1-9} 
t b & \c 5.6 pb    & \c 182974  & \c 16   &\c 3 &\c 0  &
\c0 
& \c 0  &    \\
\cline{1-9}
\c t($\to \ell \,\nu_{\ell}\,\rm b$) b W ($\to$ had)
$\ell = e, \mu, \tau$& 
\c 17.1 pb  & \c 499383 & \c 125 & \c 65 & \c 11 & \c 0 & \c 0 &
\\
\cline{1-9}
\c t($\to$ had) b W ($\to \ell \,\nu_{\ell}$) $\ell = e, \mu, \tau$& 
\c 17.1 pb  & \c 665454  & \c 229 & \c 127  & \c 14 & \c 3 & \c 0 &
\\
\cline{1-9}
\c Total Background& \multicolumn{7}{c|}{} &  0.11 fb     \\
\cline{1-9}
\end{tabular}
\caption[]{Event summary for the signal and the backgrounds after individual
cuts as 
described in the text for LHC14. In the second column, the Leading Order (LO)
cross-sections 
have been obtained using PYTHIA and ALPGEN. While calculating the final crosssection 
after Cut-V, the NLO cross-sections for signal(from PROSPINO) and appropriate K
factors 
(whenever available) for the backgrounds, as mentioned in the text, have been used. The b-tagging
efficiency has 
also been multiplied. K and L in the third column stand for $10^3$ and $10^5$
respectively. Note that the number of simulated events is always more than the 
expected number of events at LHC (at 15 fb$^{-1}$) for both the signal and backgrounds.}
\label{tablenew}
\end{table}
We also impose a b-jet reconstruction efficiency of $70\%$ \cite{b-tag:eff} 
in our analysis. We then require the following pre-selection cuts \cite{Guchait:2011fb,Ghosh:2012mc,Ghosh:2012dh}:

\begin{itemize}

\item Cut-I : We require the mass of $\rm j^{\rm higgs}$ to be in the window [119,
129]. 

\item Cut-II : Exactly one isolated lepton($\ell$) with p$_{\rm T}(\ell) >
20$GeV and no 
isolated lepton with 10 GeV $<$ p$_{\rm T}(\ell)$ $<$ 20 GeV.

\item Cut-III : The transverse mass of the lepton($\ell$)-missing P$_{\rm T}$
system 
M$^{(\ell \, \rm P_{\rm T} \hspace{-4mm} / \hspace{2mm})}_{\rm T}$ $>$ 90 GeV. 

\item Cut-IV : At this stage we again construct ``normal`` jets using the CA
algorithm 
with R parameter of 0.5, $|\eta|$ $<$ 2.5, P$_{\rm T} >$ 50 GeV. We then
calculate 
H$_{\rm T}$, which is defined as the scalar sum of the P$_{\rm T}$'s of 
all these ''normal'' jets. We define R$^{\rm b\bar{\rm b}}_{\rm T}$ = 
$\dfrac{ \rm p_{{\rm T} \, \rm b_1} + \rm p_{\rm T \, \rm b_2}}{\rm H_{\rm
T}}$. 
Remember that $p_{{\rm T} \, \rm b_1}$ and $\rm p_{\rm T \, \rm b_2}$ are the
transverse 
momenta of the two subjets $\rm j^{\rm filt}_1$ and $\rm j^{\rm filt}_2$ which
are by now 
identified as two b-jets. We demand R$^{\rm b\bar{\rm b}}_{\rm T}$ $>$ 0.9.

\item Cut-V : Events are selected with $\rm P_{\rm T} \hspace{-5mm} /$
\hspace{1mm} 
$>$  125 GeV.

\end{itemize}

In Table-\ref{table2} and Table-\ref{tablenew} we show the signal and backgrounds 
after each selection cut for 8TeV and 14TeV center of mass energies respectively. 
For all the signal points we use the NLO cross-section from 
Prospino\cite{Beenakker:1996ed}. We simulate the t$\bar{\rm t}$, Wh, WZ and Zh
backgrounds 
using PYTHIA\cite{Sjostrand:2006za}. For the Wbb, Zbb and single top backgrounds
we 
generate the unweighted event files in ALPGEN \cite{Mangano:2002ea} and then use
the 
ALPGEN-PYTHIA interface (including matching of the matrix element hard partons and shower 
generated jets, following the MLM prescription\cite{MLM}) to perform the
showering and 
implement our event selection cuts. For the t$\bar{\rm t}$ background a K factor of 2 has been
used. 
We use the CTEQ6L parton distribution functions and set top mass 
at 172.9 GeV in our analysis. We have also used the new LHC PYTHIA 6.4 Tune Z2*
for 
the correct description of the Underlying events\cite{Tune:Z2*}. We see that a 
$S/\sqrt{B}$ about 4-6 can be obtained at 8TeV LHC with an integrated luminosity 
of about 30 fb$^{-1}$. At the 14TeV LHC the siuation is much better and even with 15 
fb$^{-1}$ we get fairly good number of events, while the backgrounds are totally 
under control with our selection cuts. In Table-\ref{table2} (Table-\ref{tablenew}) 
we have simulated at least 30 fb$^{-1}$ (15 fb$^{-1}$) of events for both the signal points as well as the backgrounds. 
Note that detector and other experimental effects will degrade the signal significance somewhat. We have checked that the numbers change by $\sim$10\% 
if a Gaussian smearing is added to the transverse momenta of the jets and the 
missing transverse momentum, though a faithful quantification of the detector 
effects is beyond the scope of this work.

As the masses of $\tilde\chi_1^\pm$ and $\tilde\chi_2^0$ increase (see
Table.\ref{table1}), 
their production cross-section gradually decreases. This tends to reduce the
signal 
to background ratio. But on the other hand from the benchmark-1 to benchmark-6
the mass 
difference between $\tilde\chi_2^0$ and $\tilde\chi_1^0$ also increases. This
makes the 
Higgs boson more boosted thereby increasing the efficiency of the jet
substructure 
algorithm. With the increasing mass of $\tilde\chi_2^0$ in Table.\ref{table1}
these two opposite 
effects keep competing with each other. This is why the signal efficiency
initially 
increases with increasing $\tilde\chi_2^0$ mass but again starts falling down
because of 
the rapid decrease in the crosssection.  
\section{Conclusion}
\label{conclusion}

In this paper we have implemented magic fields as messengers of SUSY breaking in GMSB. 
One of the advantages of using magic fields as messengers over other generalized messengers is 
that achievement of unification is independent of the masses of
the magic messengers.

In our model the gaugino sector is parameterized by only two independent parameters, 
one of them can be taken to be the U(1) gaugino mass and the second one being
the $\tilde\zeta$ 
parameter (\Eqs{zeta}{e:zeta.tilde}). The $\zeta$ parameter can be tuned
to get various hierarchies 
among the gaugino masses which can lead to distinct phenomenological consequences. 

The sfermion sector can also be characterized by only two independent quantities. These 
are the U(1) A-parameter $A_1$ (Eq. \ref{A-parameter}) and the parameter $\eta$ (Eq. \ref{eta}). 
Choosing $\eta$, different hierarchies between the squark and slepton masses can be achieved. 
When the value of $\eta$ is close to its upper limit $\eta_0$, the
squark and Higgs masses 
at the messenger scale tend to be almost same. This allows one to have small
$\mu$ 
parameter and Higgsino-like NLSP similar to the models of EOGM with large doublet-triple splitting.

We focus on the region of parameter space where a comparatively larger splitting (about 1:6) between 
the U(1) and SU(2) gaugino masses is achieved along with the lightest supersymmetric Higgs boson mass 
about 125 GeV. We consider the direct electroweak production of $\chi_1^{\pm}$ and $\chi_2^0$ with 
$\chi_1^{\pm}$ decaying to the lightest neutralino $\chi_1^0$ and a $W$ boson,
and  $\chi_2^0$ decaying 
to $\chi_1^0$ and the lightest Higgs $h$. Because of the large splitting between
$\chi_2^0$ and 
$\chi_1^0$, the produced Higgs boson is expected to have quite large transverse 
momentum. Motivated by 
this we have analyzed the  $\ell$ + $b$\phantom{q}$\bar{b}$ + $P_T{\hspace{-4mm}
/}$~~ channel using the jet substructure 
technique. We have simulated all possible backgrounds for this final state and conclude that 
while $S/\sqrt{B}$ $\sim$ 4 -- 6 is viable (for the mass ranges of charginos and neutralinos we have 
considered) at 8 TeV LHC with a integrated luminosity of 30 fb$^{-1}$, LHC14 can do much better and even with 
15 fb$^{-1}$ of data a decent number of signal events over the backgrounds is expected. 
In our analysis, we have not considered any detector effect which is expected to degrade the signal significance to some extent.

A detailed exploration of other phenomenological consequences of this class of models including 
constraints from Flavor physics as well as other low energy experiments should be carried out and 
we plan to perform such a study in a future publication \cite{future}.   


\section{Acknowledgement}
PB would like to thank Prof. Palash B. Pal for encouragement and Diego Marques
for some comments. PB is also  grateful to Prof. Amol Dighe for his kind
hospitality at the Department of Theoretical Physics of TIFR where a large part
of this project was completed. DG thanks Prof. Monoranjan Guchait and Dr. Dipan
Sengupta for insightful discussions and technical help in the event simulation. 
DG would also like to thank Prof. Amol Dighe for his continuous support.


\end{document}